\documentclass[11pt]{article}

\usepackage[cp1251]{inputenc}

\begin{document}

\begin{center}

{\bf  V.V. Kisel, E.M. Ovsiyuk, V.M. Red'kov \\[3mm]
 EXACT SOLUTIONS FOR A QUANTUM-MECHANICAL PARTICLE \\WITH SPIN 1
 IN THE EXTERNAL HOMOGENEOUS MAGNETIC FIELD
 }

\end{center}


\begin{quotation}

With the use of the general covariant matrix 10-dimensional  Petiau -- Duffin
-- Kemmer formalism in cylindrical coordinates and tetrad there
are constructed  exact solutions of the quantum-mechanical
equation for a particle with spin 1 in presence of an external
homogeneous magnetic field. There are separated three linearly
independent types of solutions; in each case the formula for
energy levels has been found.

\end{quotation}

\section{Introduction, setting the problem
}

The  problem of a quantum-mechanical particle in the  external homogeneous  magnetic field is well-known
in theoretical physics. In fact, only two cases are considered: a scalar (Schr\"{o}dinger's)
 non-relativistic particle with spin 0, and fermions
(non- relativistic Pauli's and relativistic Dirac's) with spin
$1/2$ (the first investigation were \cite{1,2,3,4}). In the
present paper,  exact solutions for a vector particle with spin 1
will be constructed explicitly. The most popular
quantum-mechanical problem for such a particle is that in presence
of external Coulomb potential \cite{4}.

 To treat the problem we take the matrix Petiau -- Duffin --  Kemmer approach in the theory of the vector particle
 extended to a general covariant form on the base of  tetrad formalism (recent
 consideration and list of references see in  \cite{5,6}).

The main equation in tetrad form is \cite{6}
\begin{eqnarray}
 \left [ \; i\;
\beta^{\alpha}(x) \; ( \partial_{\alpha} + B_{\alpha} - i {e \over
\hbar } A_{\alpha} ) \;- \;
 {Mc\over \hbar}\; \right ] \; \Psi (x) = 0 \; ,
\nonumber
\\
\beta^{\alpha} (x) = \beta^{a} \; e_{(a)}^{\alpha} (x) , \qquad
B_{\alpha}(x) = {1 \over 2 } \; J^{ab} \; e_{(a)}^{\beta}
\nabla_{\alpha} e_{(b)\beta} \; ; \label{2.1}
\end{eqnarray}

\noindent $e_{(a)}^{\alpha} (x)$ is a tetrad,  $J^{ab}$ stands for generators for
 10-dimensional representation of the Lorentz group referred to 4-vector and anti-symmetric tensor
  (for brevity  we note
 $Mc / \hbar$ as $M $).
To the homogeneous magnetic field  ${\bf B} = (0,0,B)$ corresponds 4-potential
\begin{eqnarray}
 A^{a} = ( \; 0, \vec{A}
\;) = (\;  0, {1 \over 2} \; \vec{B} \times \vec{r} \;) =
 {B \over 2} \; (\;  0, -x^{2}, + x^{1}, 0 \; ) \; ;
\nonumber
\end{eqnarray}

\noindent  in the cylindric coordinates it is given by a simple expression
\begin{eqnarray}
( ct , r, \phi, z )  \; , \;\; dS^{2} = c^{2} dt^{2} - d r^{2} -
r^{2} \; d\phi^{2} - dz^{2} \;, \nonumber
\\
A_{0} = 0 , \qquad A_{r} = 0 , \qquad A_{\phi} = - {B r^{2} \over
2} , \qquad A_{z} = 0 \; . \label{2.2c}
\end{eqnarray}

Choosing a diagonal cylindric tetrad
\begin{eqnarray}
e_{(0)}^{\alpha} = (1,0,0,0 ) \; , \;\;  e_{(1)}^{\alpha} =
(0,1,0,0 ) \; ,\;\; e_{(2)}^{\alpha} = (0,0, {1 \over r }, 0 ) \;
, \;\; e_{(3)}^{\alpha} = (0,0,0,1 ) \; . \label{2.3}
\end{eqnarray}

\noindent after simple calculation,  the main equation (\ref{2.1}) is reduced to the form
 \begin{eqnarray}
\left [ \; i \beta^{0} \; \partial_{0}  +  i \beta^{1} \;
\partial_{r} + i {\beta^{2}  \over  r } \;
 ( \; \partial _{\phi} +  \; {ie B \over 2 \hbar } \;  r^{2}  \; + J^{12} \;) +
i \beta^{3} \;  \partial_{z} - M \; \right ]  \Psi (t,r,\phi,z) =
0 \; . \label{2.7}
\end{eqnarray}

\noindent For brevity we will note $(eB/2  \hbar)$ as $B$. It is best  to chose the matrices  $\beta^{a}$
in the so-called cyclic form, where  the generator  $J^{12}$ has a diagonal structure. In block-form ($ 1-3-3-3$)
these matrices are
\begin{eqnarray}
\beta^{0} = \left | \begin{array}{rrrr}
 0       &   0        &  0  &  0 \\
 0  &  0       &  i  & 0  \\
  0  &   -i       &   0  & 0\\
   0  &  0       &   0  & 0
\end {array}
\right |, \qquad \beta^{i} = \left |
\begin{array}{rrrrr}
  0       &  0       &    e_{i}  & 0       \\
    0   &  0       &   0      & \tau_{i} \\
   -e_{i}^{+}  &  0       &   0      & 0       \\
   0       &  -\tau_{i}&   0      & 0
\end {array} \right | \; ,
\nonumber \label{2.9a}
\end{eqnarray}

\noindent where   $e_{i},  \; \tau_{i}$  denote
\begin{eqnarray}
e_{1} = {1 \over \sqrt{2}} ( -i, \; 0  , \; i )\; , \qquad e_{2} =
{1 \over \sqrt{2}} ( 1 , \; 0  , \;  1 )\; , \qquad e_{3} = ( 0 ,
i  , 0)\; , \; \nonumber
\\
\tau_{1} = {1 \over \sqrt{2}} \left |  \begin{array}{ccc} 0  &  1
&  0  \\ 1 &  0  &  1  \\ 0  &  1  &  0
\end{array} \right | , \qquad
\tau_{2}= {1 \over \sqrt{2}} \left |
\begin{array}{ccc} 0  &  -i  &  0  \\ i & 0  &  -i  \\ 0  &  i  &
0
\end{array} \right | , \qquad
  \tau_{3} =   \left |
\begin{array}{rrr} 1  &  0  &  0  \\ 0  &  0  &  0   \\ 0  &  0
&  -1
\end{array} \right |  =  s_{3}\; .
\nonumber
\end{eqnarray}

\noindent Entering  eq.  (\ref{2.7}) generator $J^{12}$ is given by
\begin{eqnarray}
 J^{12} =  \beta^{1} \beta^{2} -  \beta^{2} \beta^{1}
=
  -i \left | \begin{array}{cccc}
0 & 0  &  0 & 0  \\
0 &   \tau_{3}  & 0 & 0 \\
0 & 0 &  \tau_{3} & 0 \\
0 & 0 & 0 &  \tau_{3}
\end{array} \right | = -iS_{3} \; .
\nonumber
\end{eqnarray}

\section{Separation of the variables }

With the use of special substitution (it corresponds to diagonalization of the third projections of momentum $P_{3}$
and angular momentum $J_{3}$ for a particle  with spin 1)
\begin{eqnarray}
\Psi = e^{-i\epsilon t  }  e^{im\phi}  e^{ikz} \left |
\begin{array}{c}
\Phi_{0} \\
\vec{\Phi} \\
\vec{E} \\
\vec{H}
\end{array} \right |,
\left [  \epsilon \beta^{0}   + i\beta^{1}  \partial_{r}  - {
\beta^{2}  \over  r }
 (   m + Br^{2}  - S_{3} ) - k  \beta^{3}  - M  \right ]  \left | \begin{array}{c}
\Phi_{0} \\
\vec{\Phi} \\
\vec{E} \\
\vec{H}
\end{array} \right |
 = 0  .
\label{3.2}
\end{eqnarray}

\noindent  after calculations  we arrive at the radial system of ten equations
\begin{eqnarray}
-b_{m-1} \;  E_{1} -   a_{m+1} \;  E_{3}   -     ik \;  E_{2} = M
\Phi_{0} \; , \nonumber
\\
-i  b_{m-1}  \; H_{1}  + i  a_{m+1}  \;  H_{3}  +  i \epsilon  \;
E_{2}   = M  \Phi_{2}\;, \nonumber
\\
  i  a_{m}   \; H_{2}  +i \epsilon \; E_{1}   -  k\; H_{1} = M  \Phi_{1}\;,
\nonumber
\\
 -i  b_{m}  \; H_{2} + i \epsilon  \; E_{3} +  k \; H_{3} = M  \Phi_{3}\;,
\label{3.9a}
\end{eqnarray}
\begin{eqnarray}
    a_{m}  \; \Phi_{0} -i  \epsilon  \; \Phi_{1}  = M  E_{1}\;, \qquad
    -i   a_{m}   \; \Phi_{2}  + k \;\Phi_{1} = M   H_{1}\;,
\nonumber
\\
   b_{m}   \; \Phi_{0}  -i  \epsilon  \;  \Phi_{3}  = M  E_{3}\;, \qquad
i b_{m}  \;  \Phi_{2}   -   k \; \Phi_{3}= M  H_{3}\;, \nonumber
\\
-i \epsilon  \;  \Phi_{2}  -  i  k \; \Phi_{0} =M   E_{2}\;,
\qquad
  i   b_{m-1}  \; \Phi_{1} -  i  a_{m+1}  \;   \Phi_{3}
 = M   H_{2} \; ,
\label{3.9b}
\end{eqnarray}

\noindent where special abbreviations are used for first order differential operators
\begin{eqnarray}
{1 \over \sqrt{2}}  (  {d   \over d r }    +  {m + Br^{2}   \over
r } ) =    a_{m}\; ,\qquad  {1 \over \sqrt{2}} (  - {d   \over d r
}    +  {m + Br^{2}   \over r } ) =   b_{m} \; . \nonumber
\label{3.8}
\end{eqnarray}

From (\ref{3.9a}) -- (\ref{3.9b}) it follow 4 equations for the components  $\Phi_{a}$
\begin{eqnarray}
(-b_{m-1} \; a_{m}   -  a_{m+1} b_{m}   -   k ^{2}  - M^{2} )\;
\Phi_{0} -
  \epsilon  k \;  \Phi_{2}  +  i  \epsilon  \; ( \;  b_{m-1}  \Phi_{1}    +  a_{m+1}  \Phi_{3} \;  ) = 0  \; ,
\nonumber
\\
( \; -    b_{m-1}  a_{m}   -   a_{m+1}  b_{m}  +   \epsilon ^{2}
-M^{2} \; )\;  \Phi_{2} +  \epsilon k \; \Phi_{0}   -  ik \;( \;
b_{m-1} \Phi_{1}      +  a_{m+1}   \Phi_{3} \; )         = 0 \;,
\nonumber
\\
( \; -   a_{m}  b_{m-1}    +  \epsilon^{2}   - k^{2}      -   M
^{2} \;  ) \; \Phi_{1}   + a_{m}  a_{m+1}  \;   \Phi_{3}    +    i
\epsilon \;  a_{m}  \; \Phi_{0} + i k  \; a_{m}   \Phi_{2}     = 0
\; , \nonumber
\\
(\; -  b_{m} a_{m+1}  + \epsilon ^{2}    - M^{2} - k^{2} \;  )\;
\Phi_{3} +  b_{m} b_{m-1}   \Phi_{1}    +     i \epsilon \; b_{m}
\Phi_{0}      +     ik\;  b_{m}    \Phi_{2}      = 0 \; ;
\label{3.10}
\end{eqnarray}

\section{Special simple class of solutions}

There exists a simple linear condition on  4-vector
 $\Phi_{a}$, leading to a second order differential equation. Let
 it be  $\Phi_{1}=0 \; ,  \; \Phi_{3}=0$,   the system
(\ref{3.10}) gives
\begin{eqnarray}
(-b_{m-1} \; a_{m}   -  a_{m+1} b_{m}   -   k^{2}  - M^{2} )\;
\Phi_{0} -
  \epsilon  k \;  \Phi_{2}   = 0  \; ,
\nonumber
\\
( \; -    b_{m-1}  a_{m}   -   a_{m+1}  b_{m}  +   \epsilon ^{2} -
M^{2} \; )\;  \Phi_{2} +  \epsilon k \; \Phi_{0}          = 0 \;,
\nonumber
\\
    i \;  a_{m}  ( \epsilon   \; \Phi_{0} + i k     \Phi_{2}  )   = 0 \; , \qquad
     i \; b_{m} ( \epsilon \;     \Phi_{0}      +     ik\;     \Phi_{2})      = 0 \; .
\label{4.1a}
\end{eqnarray}

\noindent From two last equations in (\ref{4.1a})  we conclude that
\begin{eqnarray}
 \epsilon \;  \Phi_{0} +  k  \Phi_{2}     = 0
\label{4.1b}
\end{eqnarray}

\noindent correspondingly, the first two in (\ref{4.1a}) (\ref{4.1a})
take the form
\begin{eqnarray}
(\; -b_{m-1} \; a_{m}   -  a_{m+1} b_{m}  + \epsilon ^{2}  - k^{2}
- M^{2} \; )\;   \Phi_{0}
  = 0  \; ,
\nonumber
\\
( \; -    b_{m-1}  a_{m}   -   a_{m+1}  b_{m}  +   \epsilon ^{2} -
k ^{2}  - M^{2} \; )\;  \Phi_{2}
   = 0 \; .
\label{4.2}
\end{eqnarray}

\noindent Because, we can readily get
\begin{eqnarray}
-  b_{m-1}  a_{m} - a_{m+1}   b_{m} = {d^{2} \over dr^{2}}  + {1
\over r}{d \over d r} - {(m+Br^{2})^{2} \over r^{2}}  = \Delta \;
, \nonumber
\end{eqnarray}

\noindent eqs.  (\ref{4.2}) are differential equations of one the same type that is operative in the theory
of a scalar particle in magnetic field
\begin{eqnarray}
(  \Delta + \epsilon ^{2}  -   k^{2}  - M^{2} \; )\;   \Phi_{0}
  = 0  \; , \qquad
( \Delta   +   \epsilon ^{2}  - k ^{2} - M^{2} \; )\;  \Phi_{2}
   = 0 \; .
\label{4.3}
\end{eqnarray}

\noindent All the remaining  component of the 10-dimensional function
can be  found straightforwardly as  in accordance with the relations
\begin{eqnarray}
\Phi_{1} =0 , \; \Phi_{3}= 0\;, \qquad \epsilon \Phi_{0} + k
\Phi_{2} = 0 \; , \nonumber
\\
    a_{m}  \; \Phi_{0}    = M  E_{1}\;,\qquad
  a_{m}   \; \Phi_{2}  = iM   H_{1}\;, \qquad
   b_{m}   \; \Phi_{0}   = M  E_{3}\;,
\nonumber
\\
 b_{m}  \;  \Phi_{2}   = -i  M  H_{3}\;,
\qquad ( \epsilon  \;  \Phi_{2}  +  k \; \Phi_{0} )  = i M
E_{2}\;, \qquad
 0  =    H_{2} \; .
\label{4.4}
\end{eqnarray}

\noindent
In general, there must exist three types of solutions for the particle with spin 1, we have   found  only one that.

\section{General analysis of the radial  equations}

Eqs.  (\ref{3.10}) can be transformed to the form
\begin{eqnarray}
[\; -b_{m-1} \; a_{m}   -  a_{m+1} b_{m}   + \epsilon^{2}    -
M^{2} -   k ^{2}  \; ] \;
  ( \; k \; \Phi_{0}  + \epsilon \; \Phi_{2}\; )
 = 0 \; ,
\nonumber
\\
\; [ \; -b_{m-1} \; a_{m}   -  a_{m+1} b_{m} + \epsilon^{2} -
k^{2} - M^{2} \; ]\; ( \epsilon \;  \Phi_{0} +  k \; \Phi_{2} ) =
\nonumber
\\
= (\epsilon^{2} - k^{2}) \; [\; ( \epsilon \;  \Phi_{0} +  k \;
\Phi_{2} )
  - \; ( \; i b_{m-1}  \Phi_{1}    +  i a_{m+1}  \Phi_{3} \;  )\; ]  \;;
\label{5.2a}
\\[2mm]
( \; -   a_{m}  b_{m-1}    +  \epsilon^{2}   - k^{2}      -   M
^{2} \;  ) \; \Phi_{1}   + a_{m}  a_{m+1}  \;   \Phi_{3}    +    i
\epsilon \;  a_{m}  \; \Phi_{0} + i k  \; a_{m}   \Phi_{2}     = 0
\; , \nonumber
\\
(\; -  b_{m} a_{m+1}  + \epsilon ^{2}    - M^{2} - k^{2} \;  )\;
\Phi_{3} +  b_{m} b_{m-1}   \Phi_{1}    +     i \epsilon \; b_{m}
\Phi_{0}      +     ik\;  b_{m}    \Phi_{2}      = 0 \; .
\label{5.2b}
\end{eqnarray}

\noindent
Let us introduce new variables
\begin{eqnarray}
F(r) =   k \; \Phi_{0}(r)  + \epsilon \; \Phi_{2}(r)\;  , \qquad
 G (r) =  \epsilon \;  \Phi_{0}(r)  +  k \; \Phi_{2} (r) \; ,
 \label{D}
 \end{eqnarray}

\noindent then eqs.  (\ref{5.2a}) -- (\ref{5.2b})  read
\begin{eqnarray}
[\; -b_{m-1} \; a_{m}   -  a_{m+1} b_{m}   + \epsilon^{2}    -
M^{2} -   k ^{2}  \; ] \;F
   = 0 \; ,
\nonumber
\\
\; [ \; -b_{m-1}  a_{m}   -  a_{m+1} b_{m}    - M^{2}\;   ]\; G =
  - (\epsilon^{2} - k^{2} ) \;  (  i b_{m-1}  \Phi_{1}    +  i a_{m+1}  \Phi_{3}   )\; ]  \;,
\label{5.4a}
\end{eqnarray}
\begin{eqnarray}
( \; -   a_{m}  b_{m-1}    +  \epsilon^{2}   - k^{2}      -   M
^{2} \;  ) \; \Phi_{1}   + a_{m}  a_{m+1}  \;   \Phi_{3}    +    i
a_{m} \; G    = 0 \; , \nonumber
\\
(\; -  b_{m} a_{m+1}  + \epsilon ^{2}    - M^{2} - k^{2} \;  )\;
\Phi_{3} +  b_{m} b_{m-1}   \Phi_{1}    +     i  \; b_{m} \;  G =
0 \; . \label{5.4b}
\end{eqnarray}

For  two  equations in  (\ref{5.4b}), let us multiply the first  (from the left)  by  $b_{m-1}$ and
the second by the  $a_{m+1}$ , which result in
\begin{eqnarray}
- b_{m-1}  a_{m}  ( b_{m-1}  \Phi_{1} )     + ( \epsilon^{2}   -
k^{2}      -   M ^{2} )   ) (b_{m-1} \Phi_{1} )  + b_{m-1} a_{m}
(a_{m+1}    \Phi_{3} )   +    i  b_{m-1} a_{m}  G    = 0 \; ,
\nonumber
\\
- a_{m+1}  b_{m} (a_{m+1} \Phi_{3})  + (\epsilon ^{2}    - M^{2} -
k^{2} ) (a_{m+1} \Phi_{3}) + a_{m+1}  b_{m} (b_{m-1}   \Phi_{1}) +
i  \; a_{m+1} b_{m} \;  G     = 0 \; . \nonumber
\\
\label{5.5b}
\end{eqnarray}

\noindent Again, let us introduce  two new variables
\begin{eqnarray}
 b_{m-1}  \Phi_{1}  = Z_{1}\;  , \qquad a_{m+1}    \Phi_{3}  = Z_{3} \; ;
\label{5.6}
\end{eqnarray}

\noindent eqs.  (\ref{5.5b}) read as follows
\begin{eqnarray}
- b_{m-1}  a_{m}  Z_{1}     + ( \epsilon^{2}   - k^{2}      - M
^{2}    ) Z_{1}  + b_{m-1} a_{m}  Z_{3}    +    i  b_{m-1} a_{m}
G    = 0 \; , \nonumber
\\
- a_{m+1}  b_{m}  Z_{3}   + (\epsilon ^{2}    - M^{2} - k^{2} )
Z_{3} + a_{m+1}  b_{m}  Z_{1}     +     i  \; a_{m+1} b_{m} \;  G
= 0 \; . \label{5.7}
\end{eqnarray}

\noindent With the aid of new  functions  $f(r), g(r)$
\begin{eqnarray}
Z_{1} = {f+g\over 2}\; , \;\; Z_{3}= {f-g \over 2} \; , \qquad
Z_{1}+ Z_{3} = f\; ,  \;\;  Z_{1}- Z_{3} = g \; ; \label{5.8}
\end{eqnarray}

\noindent the system (\ref{5.7})  is transformed to the following ones
\begin{eqnarray}
- b_{m-1}  a_{m} \;  g     +
 ( \epsilon^{2}   - k^{2}      -   M ^{2}    ) {f+g\over 2}
   +    i  b_{m-1} a_{m}  G    = 0 \; ,
\nonumber
\\
 a_{m+1}  b_{m} \;  g   + (\epsilon ^{2}    - M^{2} - k^{2} )  {f-g \over 2}
   +     i  \; a_{m+1} b_{m} \;  G     = 0 \; .
\label{5.9}
\end{eqnarray}

\noindent Combining these equations we get
\begin{eqnarray}
[\; - b_{m-1}  a_{m}    -  a_{m+1}  b_{m}  +  \epsilon^{2}   -
k^{2}      -   M ^{2}    \; ]\;  g
      +    i (  b_{m-1} a_{m}  -   \; a_{m+1} b_{m}  ) \; G    = 0 \; ,
\nonumber
\\
( - b_{m-1}  a_{m}       + a_{m+1}  b_{m}  ) \;  g +
 ( \epsilon^{2}   - k^{2}      -   M ^{2}    ) f
   +    i  ( b_{m-1} a_{m}  +  a_{m+1} b_{m} )  G    = 0 \; .
\label{5.10}
\end{eqnarray}

In these variables,  eqs.  (\ref{5.4a})  can be written as
\begin{eqnarray}
(\; -b_{m-1} \; a_{m}   -  a_{m+1} b_{m}   + \epsilon^{2}    -
M^{2} -   k ^{2}  \; ) \;F
   = 0 \; ,
\nonumber
\\
( \; -b_{m-1}  a_{m}   -  a_{m+1} b_{m}    - M^{2}\;   )\; G =
  - i (\epsilon^{2} - k^{2} ) \;  f   \; .
\label{5.11}
\end{eqnarray}

\noindent
Further, with  the use of identities
\begin{eqnarray}
-b_{m-1} \; a_{m}   -  a_{m+1} b_{m} =  \Delta  \; , \qquad -
b_{m-1} \; a_{m}   +  a_{m+1} b_{m} = 2 B  \; . \label{5.12}
\end{eqnarray}

\noindent eqs.  (\ref{5.11})  and  (\ref{5.10})   can be written down as follows
\begin{eqnarray}
( \Delta    + \epsilon^{2}    - M^{2} -   k ^{2}   ) \;F
   = 0 \; ,
\nonumber
\\
  \Delta \;  G  =  M^{2}     G      - i  (\epsilon^{2} - k^{2} ) \; f     \; ,
\nonumber
\\[2mm]
(  \Delta   +  \epsilon^{2}   - k^{2}      -   M ^{2}     )\;  g =
           2i B  \; G     \; ,
\nonumber
\\
 ( \epsilon^{2}   - k^{2}      -   M ^{2}    ) \; f
   -     i \Delta \; G   +   2B \;     g   = 0 \; .
\label{5.13}
\end{eqnarray}

\noindent
With the help of the second equation, from the forth one it follows  the linear relationship
\begin{eqnarray}
              \; f =
   -     i    \;    G        +  { 2B  \over  M ^{2} } \;     g    \; .
\label{5.14}
\end{eqnarray}

\noindent Now, excluding the function  $f$ in the second one in
(\ref{5.13})
\begin{eqnarray}
 ( \Delta + \epsilon^{2} - k^{2}  -  M^{2}  ) \;    G  =     - i  (\epsilon^{2} - k^{2} )
   { 2B  \over  M ^{2} } \;     g       \;.
\label{5.15}
\end{eqnarray}

Thus, the general problem is reduced  to the system of four equations
\begin{eqnarray}
( \Delta    + \epsilon^{2}    - M^{2} -   k ^{2}   ) \;F
   = 0 \; ,
   \nonumber
   \\
              \; f =
   -     i    \;    G        +  { 2B  \over  M ^{2} } \;     g    \; ,
\nonumber
\\
( \;  \Delta   +  \epsilon^{2}   - k^{2}      -   M ^{2}  \; )\;
g =
           2i B  \; G     \; ,
           \nonumber
           \\
 (\;  \Delta + \epsilon^{2} - k^{2}  -  M^{2} \;  ) \;    G  =     - 2i B \; {\epsilon^{2} - k^{2} \over  M ^{2} }
      \;     g       \; ,
\label{5.16}
\end{eqnarray}

The structure of this system allows to separate an evident linearly independent solution
as follows
\begin{eqnarray}
f(r)=0,\qquad g(r)=0\;, \qquad H(r)=9,
\nonumber\\
F(r) \neq 0 \; , \qquad  (  \Delta  -k^{2} -M^{2}+  \epsilon ^{2}
) \; F  = 0 \; .\label{3.32}
\end{eqnarray}

\noindent corresponding functions and energy spectrum are known (also see below).
We are to solve the system of two last equations in (\ref{5.16}), in matrix form it reads
(let $\gamma = (\epsilon^{2} - k^{2}) / M ^{2} $)
\begin{eqnarray}
( \Delta    + \epsilon^{2}    - M^{2} -   k ^{2}   ) \left |
\begin{array}{c}
g (r)\\
G(r)
\end{array} \right | =
\left | \begin{array}{cc}
0  & 2iB \\
-2iB \gamma & 0
\end{array} \right |
  \left | \begin{array}{c}
g(r) \\
G(r)
\end{array} \right | \; .
\label{5.17}
\end{eqnarray}

\noindent Let us  construct transformation changing the matrix on the right to a diagonal form
\begin{eqnarray}
( \Delta    + \epsilon^{2}    - M^{2} -   k ^{2}   )  \left |
\begin{array}{c}
g '\\
G '
\end{array} \right | =
\left | \begin{array}{cc}
\lambda_{1}   & 0 \\
0   & \lambda_{2}
\end{array} \right |
  \left | \begin{array}{c}
g '\\
G'
\end{array} \right |,
\nonumber
\\
\left | \begin{array}{c}
g '\\
G'
\end{array} \right | = S \; \left | \begin{array}{c}
g \\
G
\end{array} \right | \; , \qquad
S = \left | \begin{array}{cc}
s_{11} & s_{12} \\
s_{21} & s_{22}
\end{array} \right | .
\label{5.18}
\end{eqnarray}

\noindent The problem to solve is
\begin{eqnarray}
S   \left | \begin{array}{cc}
0  & 2iB \\
-2iB \gamma & 0
\end{array} \right | S^{-1}
  = \left | \begin{array}{cc}
\lambda_{1} & 0 \\
0 & \lambda _{2}
\end{array} \right | ,
\nonumber
\end{eqnarray}

\noindent which results in two linear systems
\begin{eqnarray}
\left \{ \begin{array}{l}
- \lambda_{1} \; s_{11} - 2iB\gamma \; s_{12} = 0 \; ,\\
2iB \; s_{11} - \lambda_{1}\; s_{12} = 0 \; ,
\end{array} \right. \qquad
\left \{ \begin{array}{l}
- \lambda_{2} \; s_{21} - 2iB\gamma \; s_{22} = 0 \; ,\\
2iB \; s_{21} - \lambda_{2}\; s_{22} = 0 \; .
\end{array} \right.
\nonumber
\end{eqnarray}

\noindent The values  of $\lambda_{1}$ and $ \lambda_{2}$
are given by
\begin{eqnarray}
\lambda_{1} = \pm 2B \sqrt{\gamma} \; , \qquad \lambda_{2} = \pm
2B \sqrt{\gamma} \; . \nonumber
\end{eqnarray}

\noindent The matrix $S$  must be degenerate, so we must use different
 $\lambda_{1},
\lambda_{2}$:
\begin{eqnarray}
\mbox{Variant} \;\; (A) \qquad \qquad  \lambda_{1} '= + 2B
\sqrt{\gamma} \;, \qquad \lambda_{2}' = - 2B \sqrt{\gamma} \;,
\nonumber
\\
i \; s_{11} -   \sqrt{\gamma} \; s_{12} = 0 \; , \qquad i \;
s_{21} +  \sqrt{\gamma}\; s_{22} = 0 \; ; \label{5.20a}
\nonumber
\end{eqnarray}
\noindent let it be
\begin{eqnarray}
s_{12}=1,  \; s_{22}=1 \;, \;
 s_{11} =  -i \sqrt{\gamma} \; , \;
  s_{21} = + i \sqrt{\gamma}\; ,
\qquad
  S = \left | \begin{array}{rr}
-i \; \sqrt{\gamma}   & 1  \\
+i \; \sqrt{\gamma}   & 1
\end{array} \right |.
\label{5.20b}
\end{eqnarray}
\begin{eqnarray}
\mbox{Variant} \;\;(B)  \qquad \qquad \lambda_{1}'' = - 2B
\sqrt{\gamma} = \lambda_{2}' \;, \qquad \lambda_{2}''  = + 2B
\sqrt{\gamma} = \lambda_{1}' \;, \nonumber
\\
i \; s_{11} +  \sqrt{\gamma} \; s_{12} = 0 \; , \qquad i \; s_{21}
-  \sqrt{\gamma} \; s_{22} = 0 \; ;
\nonumber
\label{5.21a}
\end{eqnarray}
\noindent let it be
\begin{eqnarray}
s_{12}=1,  \; s_{22}=1 \; , \;  s_{11} =  +i \sqrt{\gamma} \; , \;
  s_{21} = - i \sqrt{\gamma}\; ,
\qquad
  S = \left | \begin{array}{rr}
+i \; \sqrt{\gamma}   & 1  \\
-i \; \sqrt{\gamma}   & 1
\end{array} \right |.
\label{5.21b}
\end{eqnarray}

In the new (primed) basis, eqs.  (\ref{5.17})  take the form of two separated differential equations
\begin{eqnarray}
 (A) \;\; \left (  \; \Delta   +  \epsilon^{2}   - k^{2}      -   M ^{2}  -  2B \; \sqrt{\gamma}   \;  \right  )  \; g' =0 \; ,
\nonumber
\\
   \left ( \;  \Delta   +  \epsilon^{2}   - k^{2}      -   M ^{2}  + 2B \; \sqrt{\gamma}  \;  \right  )  \; G' =0 \; ;
\label{5.22a}
\end{eqnarray}
\begin{eqnarray}
(B) \;\; \left  ( \;  \Delta   +  \epsilon^{2}   - k^{2}      -
M ^{2} +  2B \; \sqrt{\gamma}  \;  \right  )  \; g'' =0 \; ,
\nonumber
\\
\left  ( \;  \Delta   +  \epsilon^{2}   - k^{2}      -   M ^{2} -
2B\;  \sqrt{\gamma}  \; \right   )  \; G'' =0 \; . \label{5.22b}
\end{eqnarray}

Recalling the meaning of $\Delta$, let us detail the second order equation
\begin{eqnarray}
   \left (
   {d^{2} \over dr^{2}}  + {1 \over r}{d \over d r} -
{(m+Br^{2})^{2} \over r^{2}}     + \lambda^{2}  \right  )
\varphi  (r) =0 \; , \nonumber
\\
\lambda^{2} =  \epsilon^{2}   - k^{2}      -   M ^{2}  \pm
 2B \; \sqrt{\gamma}    , \qquad \sqrt{\gamma} =  {\sqrt{\epsilon^{2} - k^{2} }\over  M  } \; .
\label{dif}
\end{eqnarray}

It is convenient to introduce a new variable $ x = B r^{2}$, then eq. (\ref{dif}) reads
\footnote{For definiteness let us consider $B$ to be positive, which does not affect generality of the analysis.
So, to infinite values of  $r$ corresponds infinite and positive values of  $x$.}
\begin{eqnarray}
x{d^{2}\varphi\over dx^{2}}+ {d\varphi\over dx}-\left(
{m^{2}\over4 x}+{x\over 4}+{m\over 2}-{\lambda^{2}\over
4B}\right)\varphi=0\,. \label{5.23}
\end{eqnarray}

\noindent With the substitution $ \varphi (x) = x^{A} e^{-Cx} f (x)
\; , $  for $f(x)$ we get
\begin{eqnarray}
x{d^{2}f\over dx^{2}}\,+\left(2A+1-2Cx\right){df\over dx}+
 \left[{A^{2}-m^{2}/4\over x}+ (C^{2}-{1\over 4} )x-2AC-C-{m\over 2}+{\lambda^{2}\over 4B}\right]f=0\,.
\nonumber \label{5.24}
\end{eqnarray}

\noindent When  $A, C$ are taken as
$ A= +  \mid m \mid /2 \; , \;  C= + 1 /2 $
the previous equation becomes simpler
\begin{eqnarray}
x{d^{2}R\over dx^{2}}\,+\left(2A+1-x\right){dR\over dx}-
\left(A+{1\over 2}+{m\over 2}-{\lambda^{2}\over 4B}\right)R=0\, ,
\nonumber
\end{eqnarray}

\noindent which is of (degenerate) hypergeometric type
\begin{eqnarray}
x \;  Y '' +( \gamma -x) Y'  - \alpha Y =0 \; , \qquad \alpha=
{\mid m \mid \over 2} +{1\over 2}+{m\over 2}-{\lambda^{2}\over
4B}\,,  \qquad \gamma= \mid m \mid +1\,. \nonumber
\end{eqnarray}

\noindent To obtain polynomials we must impose  additional condition
$ \alpha = - n\;; $ which leads to the following quantization
for
$\lambda^{2}$
\begin{eqnarray}
\lambda^{2} = 4B \; ( n + {1 \over 2} + { \mid m \mid + m \over 2}
)\; . \label{5.25}
\end{eqnarray}

Taking into account (\ref{5.22a}) -- (\ref{5.22b}), we have relations
\begin{eqnarray}
(A) \;
  (   \Delta   +  (\epsilon^{2}   - k^{2})      -   M ^{2}  -  2B  {\sqrt{\epsilon^{2} - k^{2} }\over M}      )   g' =0 \; ,
\; \sqrt{\epsilon^{2} - k^{2}}  =  { +B  +   \sqrt{B^{2} +M^{2}
(M^{2} + \lambda^{2})} \over M} \; ,
 \nonumber
\\
   (   \Delta   +  (\epsilon^{2}   - k^{2} )     -   M ^{2}  + 2B  {\sqrt{\epsilon^{2} - k^{2} }\over M}      )   G' =0 \; ;
\; \sqrt{\epsilon^{2} - k^{2}}  =  { -B  +   \sqrt{B^{2} +M^{2}
(M^{2} + \lambda^{2})} \over M} \; ; \nonumber \label{5.26}
\end{eqnarray}
\begin{eqnarray}
(B)\;
 (  \Delta   +  (\epsilon^{2}   - k^{2})      -   M ^{2}  +  2B  {\sqrt{\epsilon^{2} - k^{2} }\over M}
      )   g'' =0 \; ,
\; \sqrt{\epsilon^{2} - k^{2}}  =  { -B  +   \sqrt{B^{2} +M^{2}
(M^{2} + \lambda^{2})} \over M} \; ,
 \nonumber
 \\
   (   \Delta   +  (\epsilon^{2}   - k^{2} )     -   M ^{2}  - 2B \; {\sqrt{\epsilon^{2} - k^{2} }\over M}
        )   G'' =0 \; ;
\sqrt{\epsilon^{2} - k^{2}}  =  { +B  +   \sqrt{B^{2} +M^{2}
(M^{2} + \lambda^{2})} \over M} \; . \nonumber \label{5.27}
\end{eqnarray}

\noindent In fact, here there exist only two different possibilities
(and correspondingly two formulae for energy spectrum) :
\begin{eqnarray}
 \sqrt{\epsilon^{2} - k^{2}}  =  { +B  +   \sqrt{B^{2} +M^{2} (M^{2} + \lambda^{2})} \over M}  \; ,\qquad
 q'(r) \neq 0, \; G' = 0 \; ;
 \nonumber
 \\
\sqrt{\epsilon^{2} - k^{2}}  =  { -B  +   \sqrt{B^{2} +M^{2}
(M^{2} + \lambda^{2})} \over M} \; , \qquad
 q'(r) = 0, \; G' \neq 0 \; .
\label{5.28}
\end{eqnarray}

\noindent In turn, energy spectrum for the case (\ref{3.32}) is given by
\begin{eqnarray}
\epsilon^{2} = M^{2} + k^{2} + \lambda ^{2} \; \label{5.29}
\end{eqnarray}

Thus,  on the base of the use of general covariant formalism in the Petiau -- Duffin -- Kemmer theory
 of  the vector particle,  exact solutions for such a particle  are constructed in presence of external homogeneous magnetic
 field. There are separated three types of linearly independent solutions, and energy spectra are found.

The authors are grateful to participant of the seminar of Laboratory of theoretical physics,
Institute of Physics, National  Academy of Sciences of Belarus, for stimulating discussion.

\end{document}